\makeatletter
\AtBeginDocument{\let\LS@rot\@undefined}
\@ifundefined{@parse@version@dash}{%
\def\@parse@version#1{\@parse@version@0#1}
\def\@parse@version@#1/#2/#3#4#5\@nil{%
\@parse@version@dash#1-#2-#3#4\@nil}
\def\@parse@version@dash#1-#2-#3#4#5\@nil{%
  \if\relax#2\relax\else#1\fi#2#3#4 }
  }{}
\makeatother

\documentclass[%
 reprint,
 amsmath,amssymb,
 aps,
]{revtex4-2}

\usepackage{graphicx}
\usepackage{dcolumn}
\usepackage{bm}
\usepackage[flushleft]{threeparttable}
\usepackage{booktabs,ragged2e}
\usepackage[ruled,vlined]{algorithm2e}
\usepackage{paralist,listings}
\usepackage{amsthm}
\usepackage{amsmath, upgreek, amsfonts}
\usepackage{pgf}
\usepackage{pdflscape}
\usepackage{pdfpages}
\usepackage{tikz}
\usepackage{algorithmic}
\usepackage[autostyle, english = american]{csquotes}
\MakeOuterQuote{"}

\def\vecc#1{[#1]}
\def\idx#1{[#1]}
\def\out#1{\hat{#1}}

\newcommand{\x}{\mathbf{x}}

\newcommand{\src}{w}

\newcommand{\rdom}{\partial\mathbb{D}}

\newcommand{\ntau}{n_\tau}  
\newcommand{\nr}{n_r}  

\newcommand{\Z}{H}  
\newcommand{\X}{X}  
\newcommand{\Zg}{C}  
\newcommand{\Zs}{N}  
\newcommand{\nZg}{l}  
\newcommand{\nZs}{m}  

\newcommand{\enc}{\mathtt{Enc}}  
\newcommand{\dec}{\mathtt{Dec}}  
\newcommand{\decone}{\mathtt{Fuse}}  
\newcommand{\encg}{\mathtt{CEnc}}  
\newcommand{\encgone}{\mathtt{CEnc}_1}  
\newcommand{\encgtwo}{\mathtt{CEnc}_2}  
\newcommand{\encs}{\mathtt{NEnc}}  

\newcommand\rev[1]{\textcolor{black}{#1}}

\begin{document}

\preprint{APS/123-QED}

\title{Redatuming physical systems using symmetric autoencoders}

\author{Pawan Bharadwaj}
 \altaffiliation[Formerly at ]{Massachusetts Institute of Technology, USA.}
\affiliation{
  Indian Institute of Science, CV Raman Rd, Bengaluru, Karnataka 560012, India.
}%

\author{Matthew Li} 
\author{Laurent Demanet}%
\affiliation{%
Massachusetts Institute of Technology, 77 Massachusetts Avenue, Cambridge, MA 02139, USA.
}%




\date{\today}

\begin{abstract}
This paper considers physical systems described by hidden states and indirectly observed through repeated measurements corrupted by unmodeled nuisance parameters. 
A network-based representation learns to disentangle the coherent information (relative to the state) from the incoherent nuisance information (relative to the sensing). 
Instead of physical models, the representation uses symmetry and stochastic regularization to inform an autoencoder architecture called SymAE.
It enables redatuming, i.e., creating virtual data instances where the nuisances are uniformized across measurements.

  %
\end{abstract}

\maketitle


\paragraph*{Introduction.}

In contemporary sciences, there is increasing reliance on experimental designs involving measurements that are corrupted by unmodelled and uncontrollable \emph{nuisance variations}. 
For example, in geophysics, specifically passive time-lapse seismic monitoring, the recorded seismic data are generated by uncontrollable sources related to
tectonic stress changes in the subsurface~\cite{aki2002quantitative} 
or ocean-wave activity~\cite{Montagner2020}. 
Similarly, in astronomy, the fluorescent emissions which characterize the lunar surface's chemical composition fluctuate depending on unmodelled solar flares~\cite{tandberg1988physics,narendranath2011lunar}.  
Such designs introduce ambiguity when determining whether changes in data represent \emph{coherent information}, i.e., a signal indicating changes related to the underlying physical state, or conversely represent \emph{nuisance information}, i.e., incidental variations due to noise inherent to the data acquisition.  
Nevertheless, these uncontrollable experiments (along with others listed in Tab.~\ref{tab:expt}) remain the only feasible avenue to measure and study certain physical phenomena. This motivates the development of algorithmic tools to reliably disentangle the nuisance noise from coherent signals in these settings.

\begin{figure}
\centering
\includegraphics[width=\columnwidth]{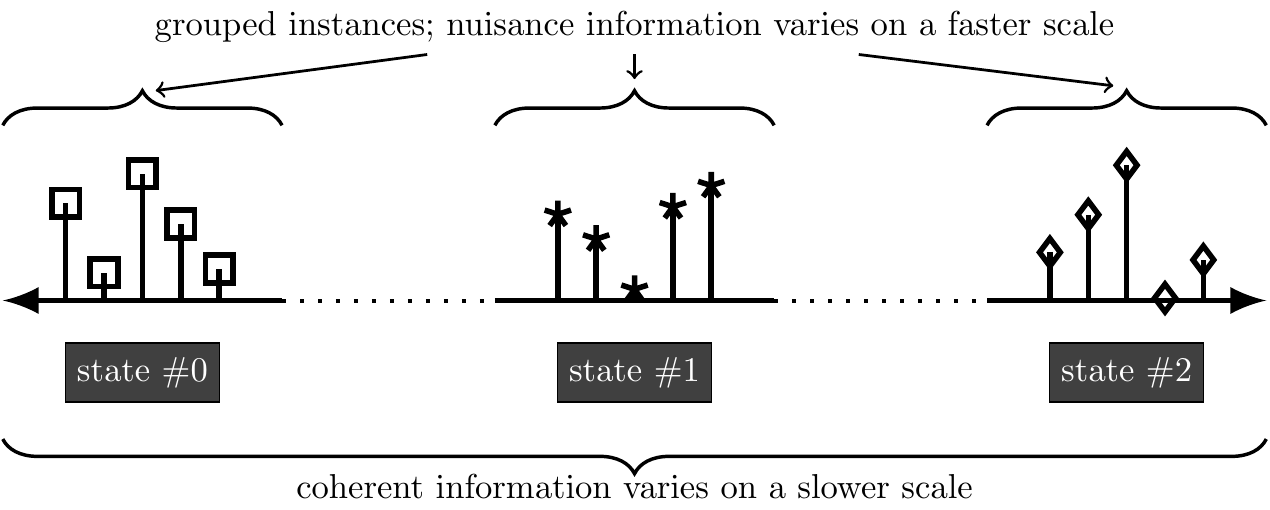}
\caption{\label{fig:slow_fast}
\rev{
Illustration of separation between two scales.
The coherent information 
(marker shape) 
related to the state varies on a much slower scale compared to the nuisance information (marker distance from the baseline).
Here, each marker represents a measurement.
}
}
\end{figure}

\rev{
This paper focuses on a sub-class of such experiments where variations in the measurements occur across two different scales:
one in which the physical processes affecting the coherent information occur, and 
the other scale corresponding to noise processes that induce variations in the nuisance information.
We assume there exists a strong scale separation in which the noise processes occur at a significantly faster rate than the former.
This assumption is crucial as it enables us to neglect variations in the coherent information within a collection of closely-spaced measurements. We refer to these \rev{closely-spaced (in scale) or} repeated measurements as \emph{instances}, and groups of instances altogether describe a \emph{(physical) state}; the dichotomy in scaling and its relation to instances and states is illustrated in Fig.~\ref{fig:slow_fast}.
}

\rev{
We further assume that the experiments we consider abundantly produce measurements describing the same physical state, albeit with different nuisance variations. Examples of experiments that satisfy these conditions are listed in Tab.~\ref{tab:expt}. 
Critically, in principle, having access to a sufficiently dissimilar collection of instances enables disentanglement of coherent information from nuisance information without any reference to the underlying physical model~\cite{pmlr-v97-locatello19a}. 
}

Our proposed approach to disentanglement decomposes each measurement into separate \emph{latent codes} which are correlated, but typically not equivalent, to the corresponding (unknown) parametric representations underlying each source of information. 
These coherent and nuisance latent codes are determined from an auto-encoding architecture with an encoder,  which maps into the latent space, and a decoder, which reconstructs the data in a near-lossless fashion. 
Notably, this avoids explicit modeling of both the physics and the nuisances by instead relying on data to inform these properties.
This framework can be seen as a vast generalization of multichannel blind deconvolution~\cite{476442}, where neural networks replace the convolutional signal model, the coherent information replaces the unknown source, and the nuisance variations correspond to the unknown filters.


\begin{table*}
\caption{
\label{tab:expt}
A representative list of experiments, where numerous instances are measured at each state that describes the physical phenomenon of interest. 
The instances exhibit dissimilarities due to the nuisance variations while sharing coherent information about the state.
}
\begin{ruledtabular}
  \begin{tabular}{cccc}
     & &\multicolumn{2}{c}{
\begin{tabular}{@{}c@{}}\emph{components of information}\\\emph{in each instance}\end{tabular}
      }\\ \begin{tabular}{@{}c@{}}\emph{experiment; goal is to} \\\emph{characterize the variation of $\cdots$} \end{tabular}
        &
\begin{tabular}{@{}c@{}}\emph{measured instances} \\ \emph{in each state comprise $\cdots$}\end{tabular}
&\begin{tabular}{@{}c@{}}\emph{coherent information} \\ \emph{describing the state }\\ \end{tabular}
&\begin{tabular}{@{}c@{}}\emph{nuisance} \\ \emph{information}\end{tabular}\\ 
\hline
     \begin{tabular}{@{}c@{}}seismic time-lapse imaging;\\ subsurface with time~\cite{verdon2010passive, kamei2014passive}\end{tabular} & \begin{tabular}{@{}c@{}}waves propagating through \\ a given subsurface region, \\ but generated by different \\ uncontrollable sources\end{tabular} & \begin{tabular}{@{}c@{}}mechanical properties \\of the subsurface region\end{tabular} & \begin{tabular}{@{}c@{}}source signature, location \\and mechanism \end{tabular} \\
\hline
     \begin{tabular}{@{}c@{}}seismology; \\ source mechanism among \\ different earthquakes~\cite{aki1972scaling,shearer2006comprehensive} \end{tabular} & \begin{tabular}{@{}c@{}}waves from an earthquake \\ measured at different receivers\end{tabular} & \begin{tabular}{@{}c@{}}earthquake's spectrum, \\geology near \\ epicenter region, etc. \end{tabular} &  \begin{tabular}{@{}c@{}} multipathing in the \\ subsurface, Doppler effects \\ from rupture propagation, etc. \end{tabular}  \\
\hline
     \begin{tabular}{@{}c@{}}lunar X-ray \\ fluorescence spectroscopy; \\lunar geology with location~\cite{clark1978utilization,narendranath2011lunar}\end{tabular} & \begin{tabular}{@{}c@{}}fluorescence from regions \\that share a given rock type\\under different solar conditions\end{tabular} & \begin{tabular}{@{}c@{}}elemental composition \\of the rock type\end{tabular} & \begin{tabular}{@{}c@{}}solar-flare \\ information\end{tabular}\\
\hline
     \begin{tabular}{@{}c@{}}asteroseismology~\cite{aerts2010asteroseismology,handler2012asteroseismology}; \\pulsating mechanism among \\ Rapidly oscillating Ap~\cite{samus2004vizier} \\ or Delta Scuti~\cite{campbell1900list}
      stars \end{tabular} & \begin{tabular}{@{}c@{}}brightness of a given star \\ measured in \\different temporal windows\end{tabular} & \begin{tabular}{@{}c@{}}physics that is symmetrical \\ under time translation, \\ e.g., star's internal structure, \\ kappa opacity mechanism, etc.  \end{tabular} & \begin{tabular}{@{}c@{}}short-lived excitation \\ mechanisms, e.g., \\ surface convection, \\ variable radial velocity, etc.\end{tabular} \\
\end{tabular}
\end{ruledtabular}
\end{table*}
%


Disentanglement into coherent and incoherent latent variables enables a reliable comparison of coherent information between instances collected from different states. 
However, this relegates the analysis into a latent space that is abstractly related to the physical system. 
To that end, we propose an additional mechanism, called \emph{redatuming}, which converts from latent coordinates back into the nominal data space representation with specifically chosen properties.
This involves combining coherent information from one instance with the nuisance information from a reference instance in order to synthesize a \emph{virtual} instance that is not originally measured (i.e., not present in the dataset).
The relevance of such virtual data instances is that they can be engineered to share their nuisance information with another (measured) data instance so that any remaining discrepancy can be solely explained from differences in the underlying coherent physical states. 
This entire process can
be alternatively viewed as ``swapping the physics'' between states.
We conjecture that this new type of redatuming can help rethink how to approach inverse problems with significant uncertainties in the forward model.

\begin{figure}
\centering
\includegraphics[width=0.5\textwidth]{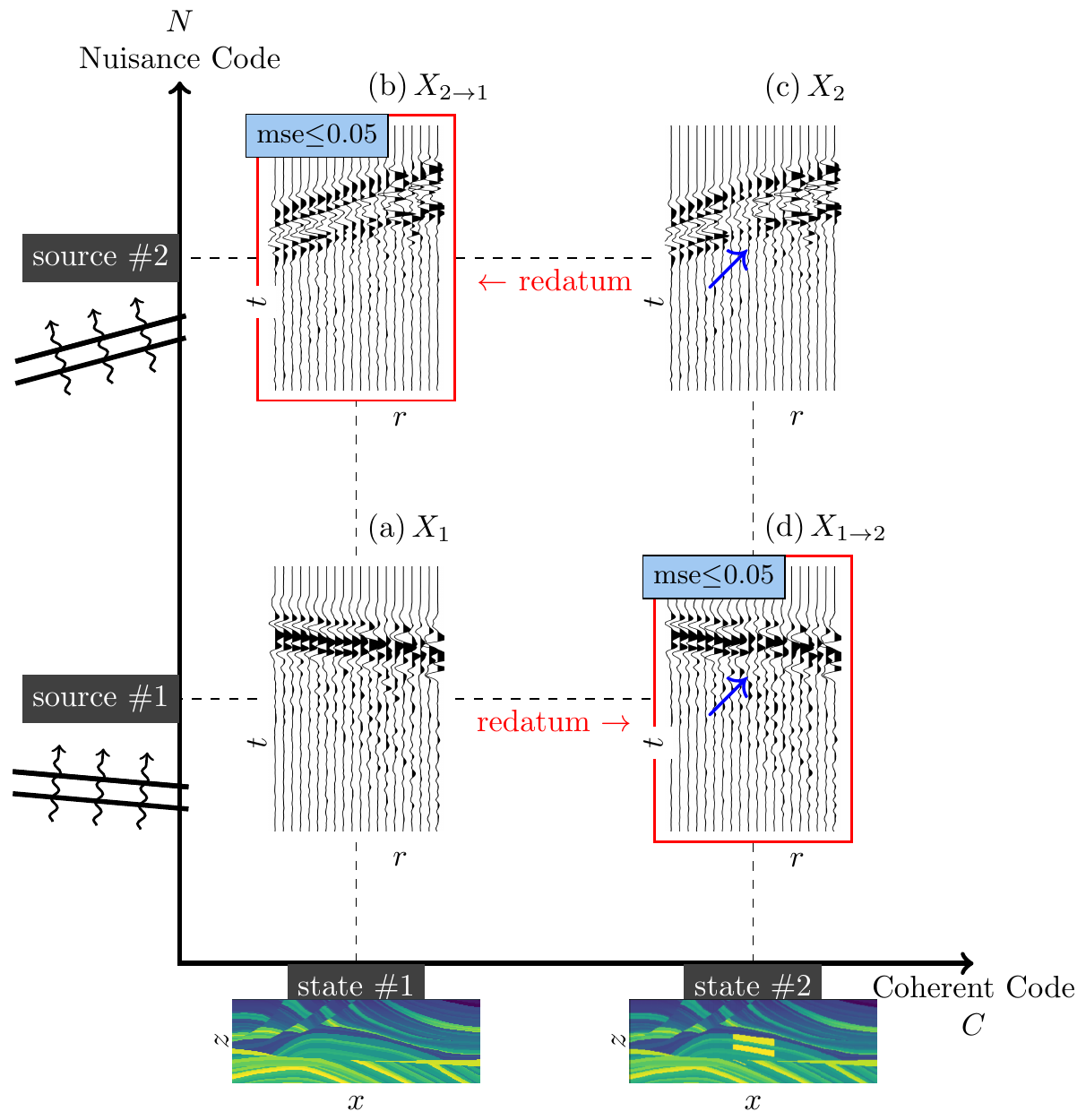}
\caption{\label{fig:intro_redatum}
Redatuming is equivalent to swapping the coherent and nuisance information in SymAE's latent space. 
Here, as SymAE learns to represent the information 
on medium (coherent) and incoming plane-wave sources (nuisance) separately, the recorded wavefield (a and c) in a seismic experiment (see Tab.~\ref{tab:expt}) can be redatumed to generate virtual measurements (b and d).
A low error confirms that the redatuming operator captures salient features of the wave-equation modeling despite a 
multipath propagation due to the complex medium inhomogeneities.
%
%
%
%
}
\end{figure}

\rev{
We illustrate redatuming using the example of time-lapse geophysical subsurface monitoring cited in Tab.~\ref{tab:expt}.
Here, seismic surveys are conducted to measure the subsurface properties (the coherent information) indirectly
by recording reflected and transmitted elastic waves generated by uncontrolled or unreliably modeled mechanisms (the nuisance information).
In this setting, it is reasonable to assume that changes to the complex heterogeneous subsurface occur on a significantly slower timescale (e.g., on the order of months) than the variations in the uncontrollable seismic sources (on the order of hours or days).
As such, the variations in the subsurface mechanical properties, as functions of the lateral scale dependent on distance $x$ and depth $z$,
can be safely neglected within each state. 
The goal is to detect and characterize changes in subsurface between the states, e.g., to distinguish between ``state \#1'' and ``state \#2'' along the abscissa of Fig.~\ref{fig:intro_redatum}~\footnote{Note that the structural complexities of the media result in multipath wave propagation, thereby complicating the inverse problem.}.
}
In each state, the measured instances constitute the time-dependent (indexed using $t$) wavefield recorded at a given set of receivers (indexed using $r$) in the medium --- plotted in Figs.~\ref{fig:intro_redatum}a and \ref{fig:intro_redatum}c.

\rev{
In effect, the seismic sources are realized with randomized signatures and locations, c.f., the nuisance variations visualized along the ordinate of Fig.~\ref{fig:intro_redatum}. This confounds 
direct visual comparison between the measured instances Fig.~\ref{fig:intro_redatum}a and \ref{fig:intro_redatum}c as it is unclear whether the localized changes (indicated by the blue arrow) are to be attributed to changes in the medium or the source.
%
Redatuming overcomes this ambiguity by generating a \emph{virtual} instance, plotted in Fig.~\ref{fig:intro_redatum}d. This virtual instance is engineered by replacing the subsurface information of Fig~\ref{fig:intro_redatum}a with that of Fig~\ref{fig:intro_redatum}c while retaining its source/nuisance information. As a result, the virtual instance can be subtracted from the reference instance (here~\ref{fig:intro_redatum}a) to qualify or quantify potential subsurface changes via standard imaging techniques.
We emphasize that redatuming enables domain experts to perform data analysis using traditional tools without any reference to the implicit latent space. 
}

These ideas are inspired by recent machine learning literature where redatuming is instead referred to as \emph{styling}, or \emph{deep-fakes}, see e.g.,~\cite{236284,10.1145/3072959.3073640,10.1145/258734.258880}, and the reliance on multiple instances is referred to as \emph{weak supervision} \cite{pmlr-v119-locatello20a,pmlr-v97-locatello19a}. 
However, we note that these communities primarily apply these tools to images with significant visual structure wherein nuisance information relates to the ``image style'', and the coherent information relates to the ``image content''.
This letter instead introduces the idea of redatuming to scientific signals, enabling us to quantify virtual-instance accuracy against explicit synthetic models rigorously. 

\paragraph*{Our Contributions.}
To achieve redatuming, we propose an unsupervised deep-learning architecture called symmetric autoencoder (SymAE). Achieving the requisite disentangled latent representation with SymAE
%
requires two deliberate architectural design choices: 
\begin{inparaenum}
  \item The encoder for the coherent latent variables is constrained to be symmetric with respect to the ordering of the instances indexed by nuisance variations.
\item The remaining latent-code dimensions are \rev{encouraged to encode independent information} by stochastic regularization that promotes dissimilarity among the instances~\footnote{In previous work,
 we used focusing constraints \cite{8680655} to maximize this dissimilarity and regularize blind deconvolution.}.
 Therefore, these remaining latent components are designed to \emph{not} represent the coherent information and correspond only to the nuisance variations.
\end{inparaenum}

\rev{
Once the coherent and nuisance information are disentangled in the latent space, redatuming is equivalent to decoding a hybrid latent code, 
specifically, a hybridization of the coherent code from one state and the nuisance code of an instance from another state.
}
We provide numerical evidence that SymAE's redatuming preserves and captures the salient features of the underlying physical modeling operator, thus enabling the use of virtual datapoints for
subsequent downstream tasks such as parameter estimation. 
\rev{
  We numerically validate that the virtual instances generated without reference to the physics 
  satisfy the governing wave equation up to a low relative mean-squared error.
This 
indicates
that SymAE redatuming is consistent with, or preserves, the physics of wave propagation.
}


The concept of redatuming appears in the context of traditional seismic inversion \cite{Wapenaar2004,schuster2006theoretical,Schuster2009,mulder2005rigorous,wapenaar2014marchenko}. The major differences with our current generalized approach, however, are: 
\begin{inparaenum}
\item the seismic-specific redatuming is limited to swapping sources or receivers from one state to another --- in contrast, SymAE aims to swap any information that is coherent across the instances;
\item \rev{seismic redatuming either requires prior knowledge about the subsurface or 
uses physics-derived relations with convolutions or cross-correlations} --- in contrast, SymAE derives the redatuming operators from the recorded data in an unsupervised manner, unlocking processing for far more general situations than cross-correlations allow.
\end{inparaenum}
We refer the reader to \cite{Mordret2014, DeRidder2014, VanderNeut2016} for examples of analytical-based redatuming applied to specific geophysical settings.

\rev{
SymAE heavily relies on imposing symmetries in the encoder to separate the latent code. 
This idea of using symmetry, or equivalently physical priors, to promote structure in the neural networks has been proposed in various works, for instance: \cite{Mattheakis2019} 
embedded even/odd symmetry of a function and energy conservation into a neural network 
by adding special \emph{hub} layers; \cite{Cohen2019} propose gauge equivariant CNN layers to capture rotational symmetry;  
\cite{Greydanus2019} structures their networks following a Hamiltonian in order to learn physically conserved quantities and symmetries. The choice of symmetry is bespoke to each application, and the identification of valid symmetries in our physical prior is one of the contributions and insights of SymAE.}

\paragraph*{Datapoints and Notation.}
In this section, we describe the training set $\{X_i\}_{i=1, \ldots, n_X}$ that SymAE encodes to produce a compressed and disentangled representation.
\rev{We re-iterate that we presume the scale separation illustrated in Fig.~\ref{fig:slow_fast} in our dataset. As such, each datapoint $X_i$ contains multiple instances that repeatedly capture the same physical state~$\epsilon_i$, but each instance may differ on account of nuisance variations.} 
We uniformly sample from $1$ to $n_\epsilon$ to generate the state labels $\{\epsilon_i\}_{i=1, \ldots, n_X}$ for our synthetic experiments --- in practice, the experimental conditions determine this sampling distribution. 
We emphasize that knowledge of the state labels is \emph{not} necessary for either training or testing since our framework is purely unsupervised. 
We index the instances in datapoint as $X_i[\tau]$ for $\tau = 1, \ldots, n_\tau$ such that $X_i = [X_i[1]; \ldots; X_i[n_\tau]]$. 
Each instance $X_i[\tau]$ is represented as $k$-dimensional vectors, and the determination of $k$ is specific to each experiment. 
\rev{
For the seismic experiment depicted in Fig~\ref{fig:intro_redatum}a, 
each instance is a source gather, where the dimension~$k$ is the product of the number of receivers and the length of the time series.
Each $\X_i$ comprises several sources that illuminate the same subsurface region.
}
In our notation, $\vecc{A;\,B}$ denotes a vertical concatenation of two vectors $A$ and $B$. 
Again, the collection of instances $\{X_i[\tau]\}_{\tau = 1, \ldots, n_\tau}$ for a fixed index~$i$ shares the same coherent information to the state $\epsilon_i$ but vary by $\tau$-specific nuisance variations. 

\paragraph*{Architecture.}
We refer the reader to \cite{doersch2016tutorial} for an accessible tutorial on autoencoders~\footnote{
We choose to use a deterministic autoencoding strategy for simplicity.
It is possible to formalize the ideas in this paper using the variational autoencoding framework.}.
Functionally, autoencoders are comprised of two components: an encoder $\enc$ that maps \rev{each} datapoint $\X_i$ into latent code $\Z_i=\enc(\X_i)$, and a decoder $\dec$ that attempts reconstruct to $\X_i$ from the code.
Traditionally, both functions $\enc$ and $\dec$ are determined by minimizing the reconstruction loss
\begin{eqnarray}
	\label{eqn:ae_loss}
	\enc,\,\dec=\operatorname*{arg\,min}_{\enc,\,\dec}  \sum_i \|\X_i-\dec(\enc(\X_i))\|^2
\end{eqnarray}
over the training dataset.
%
%
When non-linear parameterizations are used for both $\enc$ and $\dec$, the latent representation no longer describes the geometry of the datasets using linear subspaces \cite{klys2018learning}. However, this representation can efficiently compress the information \cite{Dai2019a}.

SymAE builds on non-linear autoencoders but requires additional modifications as a direct application of traditional autoencoding ideas will not ensure that the coherent and nuisance information are encoded into separate components (dimensions) in the latent space. 
%
%
To achieve this separation, SymAE relies on the unique encoder structure as depicted in Fig.~\ref{fig:network}~\footnote{\rev{We provide Tensorflow-style~\cite{Abadi2016} algorithms in the supplementary material that detail the implementation of SymAE. Additionally, following the double-blind review process, a link to a Github repository containing reproducible code will be made available in this footnote.}}.
The encoder structure can be mathematically described by
\begin{eqnarray}
\enc(&\X_i&) = \nonumber \\
\vecc{\encg(&\X_i&);\,\encs(\X_i\idx{1});\,\ldots;\,\encs(\X_i\idx{\ntau})}.
\end{eqnarray}
This output corresponds to a latent code $\Z_i$ which is partitioned into interpretable components.
Specifically, each datapoint $\X_i = [X_i[1]; \ldots; X_i[n_\tau]]$ is represented as 
a structured latent code $\Z_i=\vecc{\Zg_i;\,\Zs_i\idx{1};\,\ldots;\,\Zs_i\idx{\ntau}}$ in which
the sub-components $\Zg_i = \encg(\X_i)$ contain coherent information in $\X_i$ 
while the remaining sub-components $\Zs_i\idx{\tau} =\encs(\X_i\idx{\tau})$ encode the complementary  instance-specific nuisance information.
Note the dimensions $\nZg$ and $\nZs$ of the latent codes $\Zg_i \in \mathbb{R}^{\nZg}$ and $\Zs_i[\cdot] \in \mathbb{R}^{\nZs}$ are user-specified hyperparameters which need not coincide.

%
Subsequently, 
SymAE's decoder $\decone$ non-linearly combines code
$\Zg_i$ with each instance-specific code $\Zs_i[\cdot]$ 
to reconstruct the original datapoint, instance-by-instance, viz.
\begin{eqnarray}
\label{eqn:decode_instance}
\out{\X_i}&=&\dec(\Z_i) = \dec(\vecc{\Zg_i;\,\Zs_i\idx{1};\,\ldots;\,\Zs_i\idx{\ntau}}) \nonumber\\
&=&
  \vecc{\decone(\vecc{\Zg_i;\,\Zs_i\idx{1}});\,\ldots;\,\decone(\vecc{\Zg_i;\,\Zs_i\idx{\ntau}})}.
\end{eqnarray}
We do not enforce any constraints on $\decone$ in our experiments and parametrize it with standard deep learning building blocks~\footnote{We provided more architectural details in the supplementary material}.

\begin{figure*}
\centering
\includegraphics[width=0.85\textwidth]{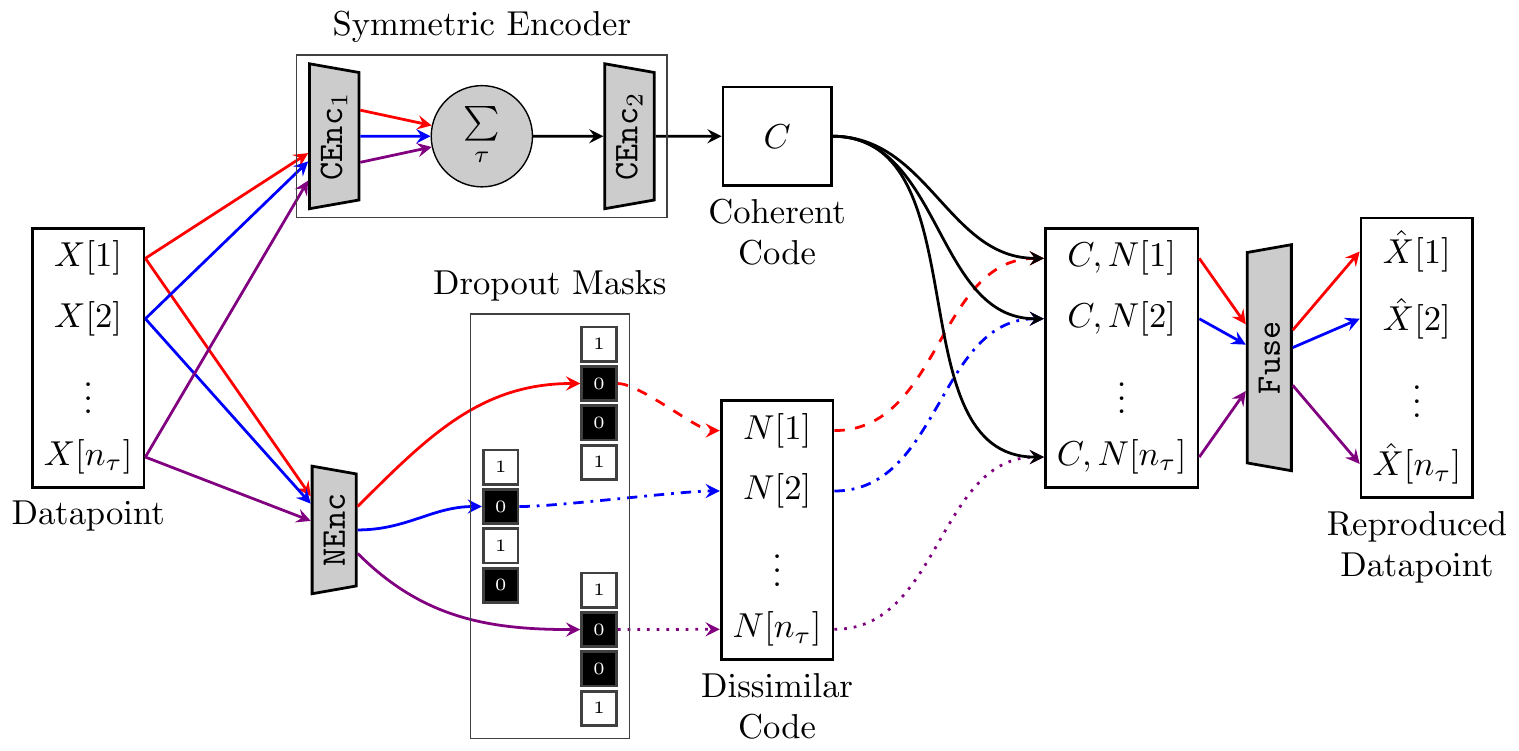}
\caption{
  \label{fig:network}
  Architecture of symmetric autoencoder.
  The information that is coherent across the instances of a datapoint can only propagate through the network via solid arrows --- notice the stochastic regularization employed to prevent its propagation. 
  We used colored arrows to indicate the propagation of the remaining instance-specific nuisance information --- notice that a symmetric function, i.e., symmetric w.r.t. the order of the instances, prevents its propagation.
  As a result, the autoencoder disentangles the coherent information from the nuisance variations in the latent space.
\rev{
We omitted the subscript $i$ for $\X$, $\Zg$ and $\Zs$.}
}
\end{figure*}

We ensure that $\encg$, the \emph{coherent encoder}, encodes at most the coherency or similarity among the instances in $\X_i$~by enforcing invariance with respect to permutations of the instances within the datapoint.
Mathematically, this condition requires that 
\begin{eqnarray}
\label{eqn:encg_sym}
\Zg_i=\encg(\X_i)=\encg(\X_i\idx{\varPi(1{:}\ntau)})
\end{eqnarray}
for all permutations $\varPi$ along the instance dimension the output.
\rev{This symmetry invokes the dichotomy of scale assumed in the data -- since only nuisance variations are assumed to vary along the ``fast scale'' $\tau$ and that any coherent changes are negligible, this permutation invariance ensures that nuisance information cannot be encoded using $\encg$ without significant loss of information. It follows from the 
pigeonhole principle that only coherent information can remain in $\Zg_i$ if a low auto-encoding loss is achieved.}

SymAE's coherent encoder explicitly achieves the invariance mentioned above using permutation-invariant network architectures following \cite{Zaheer2017} 
which provide universal approximation guarantees for symmetric functions.
\rev{
These architectures use 
pooling functions such as the $\mathrm{mean}$ or the $\mathrm{max}$ across the instances to ensure permutation invariance.
We refer to \cite{Ilse2018} for a review of alternative pooling functions, including attention-based pooling.}
In our experiments, the data due to each source instance are   
transformed using $\encgone$ and summed along the instance dimension. This output is then processed by $\encgtwo$ resulting in
\begin{eqnarray}
	\label{eqn:enc_symmetric}
\Zg_i=\encgtwo\left(
	\frac{1}{n_\tau} \sum_{\tau=1}^{n_\tau} \encgone(  \X_i\idx{\tau})
\right),
\end{eqnarray}
yielding the network architecture of $\encg$.
\rev{Intuitively, $\encgone$ extracts 
coherent information from each of the instances 
while the summation encourages them to be aligned. This information is further compressed using $\encgtwo$.}
The functions  
$\encgone$ and $\encgtwo$ 
are parametrized by compositions of  
fully connected layers and convolutional layers.  
%
We emphasize that  
the key observation in eq.~\ref{eqn:enc_symmetric} is that 
the summation 
of the \emph{transformed instances} $\encgone(\X_i\idx{\tau})$ 
is 
symmetric with respect to the ordering of instances. This 
ensures that 
the desired symmetry (eq.~\ref{eqn:encg_sym}) is achieved. 

In contrast,
the purpose of $\encs$, the \emph{nuisance encoder}, is to capture the nuisance information specific to each instance of a datapoint. Critically, we do not want the decoder $\decone$ to ignore the $\Zg_i$ component in favor of using purely $\Zs_i[\cdot]$ information for reconstruction. We desire disentanglement of the latent codes.
Whereas $\encg$ achieves this via symmetry, for the nuisance encoder this separation is encouraged through the use of \emph{stochastic regularization} viz,
\begin{eqnarray}
    \Zs_i[\tau] = \encs(\X_i[\tau]) + \textrm{``strong noise''}.
\end{eqnarray}
\rev{
Intuitively, this idea hinges on the assumption that coherent information does not vary with the ``fast scale''~$\tau$ indexing each instance. As such, obfuscating each element $\Zs_i$ via noise introduces artificial dissimilarities along this scale; this, therefore, encourages the decoder to instead rely on the coherent code (held constant for each instance, c.f. eq.~\eqref{eqn:decode_instance}) to reconstruct the coherent information. Similarly, as before, it follows from the
 pigeonhole principle that the nuisance codes $\Zs_i$ must contain at most information relevant to nuisance information if a low auto-encoding loss is achieved.}

In our experiments, we implement this noise using either Bernoulli dropout regularization~\cite{srivastava2014dropout} with probability $p$ or Gaussian dropout with unit mean and $p(1-p)$ variance \cite{wang2013fast,kingma2015variational}. 
In either case, the strength of the noise is proportional to $p$, which is a hyperparameter the user must tune.
Critically, however, each $\Zs_i$ must still be expressive enough to encode nuisance-specific information. 
The balance between regularization strength~$p$ and the dimension (i.e., expressivity) of the latent codes is user-determined on an external validation set. 
The SymAE components $\encs$, $\encg$ and 
$\decone$ are trained concurrently by minimizing Eq.~\ref{eqn:ae_loss} with the regularization mechanism just described. 
\rev{
We emphasize that the stochastic regularization is not employed to reduce over-fitting and improve generalization error in the conventional sense, see, e.g., \cite{labach1904survey} for a survey on stochastic techniques used in neural network training. Instead, the intention is to promote learning dissimilar representations across nuisance codes.
At test-time, the entirety of the $\Zs_i$ code is sent unaltered and unobfuscated into the decoder. 
}


%
%
Finally, note that we only constrained the encoders to avoid "cross-talk" while disentangling the coherent and nuisance information. 
Implicitly the success of SymAE, therefore, requires a sufficiently large number of instances with \emph{dissimilar} nuisance variations in order to achieve the desired structure of the latent space. 
We leave an examination of characterizations of physical models which are amenable to disentanglement to future work.

%


%

%
%
\paragraph*{Redatuming into Virtual Instances.}
A trained SymAE learns a representation with disentangled coherent and nuisance information.
Redatuming data becomes equivalent to manipulations in the latent space --- as illustrated in the Fig.~\ref{fig:intro_redatum}, where virtual instances are generated by swapping latent coordinates.  %
In general, the coherent information in the $\tau$-th instance of a datapoint $\X_i$ can be swapped with that of another datapoint $\X_j$ using 
\begin{eqnarray}
  \label{eqn:virtual}
	\out{\X}_{i\rightarrow j}\idx{\tau}=\decone(\vecc{\encg(\X_j);\,\encs(\X_i\idx{\tau})}).
\end{eqnarray}
Here, $\X_j$ is an observation of a different state compared to $\X_i$. Notice that the nuisance information in the virtual datapoint $\out{\X}_{i\rightarrow j}$ is identical to that of the original datapoint $\X_i$.
Consequently, we attribute the difference between $\X_i$ and $\hat{\X}_{i\rightarrow j}$ to the changes between the physical states.
As a demonstration, the observed and virtual instances from the seismic experiment are embedded into the SymAE's latent space in Fig.~\ref{fig:intro_redatum}.
%
\paragraph*{Experiments.}
\rev{ 
We now detail the application of SymAE towards experiments that monitor 
subsurface changes using seismic waves.
As noted earlier, the measurements vary on two different (time) scales.
\begin{inparaenum}
\item The slower time scale is associated with the subsurface changes that typically occur in the order of months. 
As such, the goal is to detect or determine variations in the coherent (subsurface) information between seismic surveys (e.g., baseline and monitor).
\item The faster time is usually on the order 
of the duration of the seismic survey, i.e., either hours or days; the variation in the coherent information is negligible on this scale. 
During each survey, waves from numerous uncontrollable sources, here taken to be the nuisance information, are recorded as instances.
\end{inparaenum}
}
\rev{
For our synthetic experiments, an instance
is modeled as the pressure wavefield $u(\x,t)$ from a finite-difference solver with absorbing boundary conditions for the acoustic wave equation:
\begin{eqnarray}
  \label{eqn:wav}
  \frac{1}{c^2(\x)}\frac{\partial^2 u}{\partial t^2} - \nabla\cdot\left(\nabla u\right) = \delta(\x-\x_s)\src(t).
\end{eqnarray}
Here, $\x=\vecc{x,z}$ denotes the Cartesian coordinate vector and $t$ denotes time.
The medium is parameterized using the wave-velocity $c(\x)$.
During the forward modeling, we vary $c$ at a slower rate compared to source parameters, i.e., position $\x_s$ and signature $\src(t)$ 
that determine the nuisance variation in each modeled instance.
}

\begin{figure*}
\centering
\includegraphics[width=\textwidth]{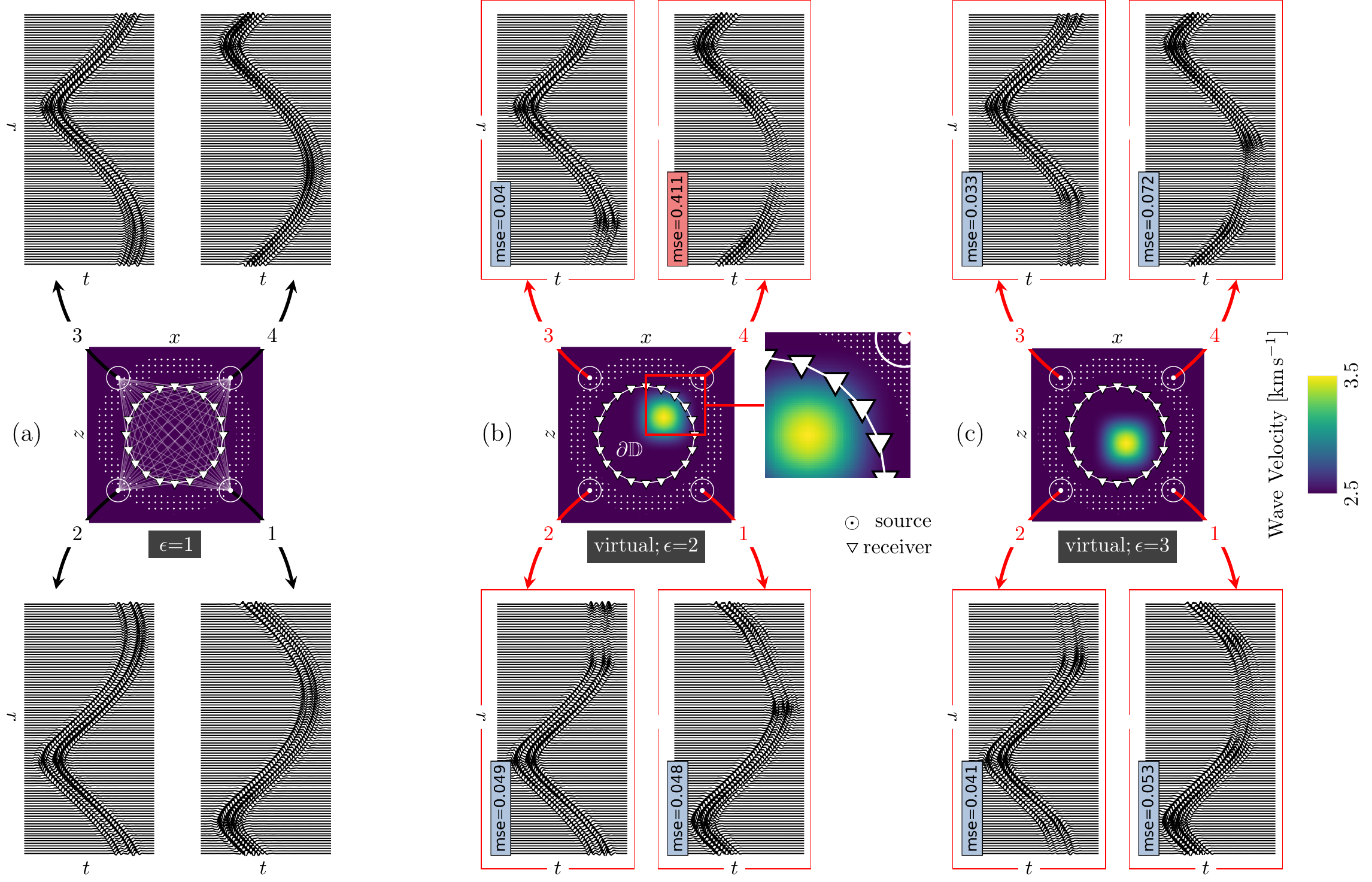}
\caption{
  \label{fig:systems_circ}
  Deep redatuming of waves recorded on $\rdom$ due to point sources in the dotted region. 
  This experiment illustrates 
  that SymAE can isolate (coherent) information on the medium (Gaussian) perturbation in its representation --- however, the perturbation has to lie within $\rdom$. 
  a) $\X_1[1{:}4]$; original instances from the first state with a homogeneous medium.
  b) $\out{\X}_{1\rightarrow 2}[1{:}4]$; virtual instances generated after swapping code 
  $\Zg_1$ of the first state with code $\Zg_2$ of the second state.
  A high MSE in $\out{\X}_{1\rightarrow 2}[4]$ indicates that $\Zg_2$ fails to represent the effects of the Gaussian perturbation entirely. 
  Note that this perturbation extends beyond $\rdom$.
  c) $\out{\X}_{1\rightarrow 3}[1{:}4]$; same as (b), except a lower MSE, means that $\Zg_3$ satisfactorily represents the medium when the perturbations lie within $\rdom$.
}
\end{figure*}

\rev{
We justify that SymAE captures the salient features of the physics of wave propagation using a simple illustration.
Consider the wave-velocity of a medium that varies in a 
$2\,\text{km}\times2\,\text{km}$ region, shown in Figs.~\ref{fig:systems_circ}a--c,
across three states
with $\epsilon_i\in\{1,2,3\}$ as described in Tab.~\ref{tab:systems_circ}.
%
%
\begin{table}[b]
\caption{\label{tab:systems_circ}%
}
\begin{ruledtabular}
\begin{tabular}{lccr}
\textrm{state $\epsilon$}&
\textrm{medium perturbation\footnotemark[1]}&
\textrm{MSE\footnotemark[2]}&
\textrm{virtual MSE\footnotemark[3]}\\
\colrule
1 & none; homogeneous & $<0.01$& -\\
2 & not entirely inside $\rdom$\footnotemark[4] & $<0.01$ &high ($>0.3$)\\
3 & inside $\rdom$\footnotemark[4] & $<0.01$ &low ($<0.1$)\\
\end{tabular}
\end{ruledtabular}
\footnotemark[1]{Gaussian perturbation}
\footnotemark[2]{
normalized mean-squared error between the true $\X$ and reconstructed $\out{\X}$ datapoints
for both training and testing}
\footnotemark[3]{between virtual and synthetic instances after redatuming}
\footnotemark[4]{receiver circle with center $\vecc{0,0}$ and radius $600\,$m}
\end{table}
Point sources at $\x_s=\vecc{R\,\cos(\theta),R\,\sin(\theta)}$ with signature $\src(t)$ are used
for modeling 
instances.
The random variables $R$ and $\theta$ are 
uniformly distributed on $[0.8,1.0]\,$km and $[0, 2\pi]$, respectively.
The random source wavelet $\src(t)$ has a duration of $0.13\,$s, is sampled from a standard normal distribution and is convolved with a 25$\,$Hz high-cut filter.
After solving eq.~\ref{eqn:wav}, 
the acoustic wavefield is sampled at 
$160$ time steps and $100$ evenly-distributed receiver locations on a circle $\rdom$ to form an instance of dimension $k$=16,000.
%
We generated 8,000 instances per state and considered a total of  6,000 datapoints (with $\ntau$=$20$) for training and testing.
It is important to note that the distribution of the forcing term in eq.~\ref{eqn:wav} is independent of the state $\epsilon$ to facilitate disentanglement.
}

\rev{
We have invariably used the same medium parameters during the forward modeling of the instances in each state. 
Therefore, we hypothesize that:
\begin{inparaenum}
\item the medium parameters $c(\x)$ characterize the coherent information represented by the code $\Zg_i$, i.e., $\encg$ encodes the information related to the entire medium in each state;
\item the forcing term $w(t)$ and the source position $\x_s$ characterize the nuisance information, represented by $\Zs_i$, of a given instance.
\end{inparaenum}
We test these hypotheses numerically and show that $\encg$ does not encode the entire medium but only a portion that is coherently illuminated by all the sources.
After redatuming, we compute relative MSE between the virtual instances (generated by deep redatuming) and synthetic instances --- 
a low MSE signifies that the virtual instance satisfies the governing wave equation in eq.~\ref{eqn:wav} with appropriate 
medium and source parameters.
We now redatum four sources, $\X_1[1{:}4]$ as in Fig.~\ref{fig:systems_circ}a, picked from the first state ($\epsilon=1$).
First, we swapped $\Zg_1$ of these measurements with $\Zg_2$ to include the physics of wave-propagation related to the Gaussian perturbation in $\epsilon=2$.
The virtual instances $\hat{\X}_{1\rightarrow 2}$ are plotted in Fig.~\ref{fig:systems_circ}b --- it can be observed that source information (position and signature) remained intact during redatuming, 
confirming that $\Zg_2$ does not represent any of the source effects.
Furthermore, notice that most of the virtual instances have low MSE, 
for example, $\hat{\X}_{1\rightarrow 2}[1{:}3]$, indicating that $\Zg_2$ captured a significant portion of the Gaussian perturbation.
However, the virtual instances with source locations close to that of $\hat{\X}_{1\rightarrow 2}[4]$, plotted in Fig.~\ref{fig:systems_circ}b, have high MSE.
What is unique about these sources?
It is evident from the ray paths in Fig.~\ref{fig:systems_circ}a that these high-MSE source locations illuminate the portion of the Gaussian perturbation outside $\rdom$.
On the other hand, the region inside $\rdom$ is coherently illuminated irrespective of the source position. 
We infer that SymAE's coherent code only represents the propagation effects of inhomogeneities inside $\rdom$.
In order to further confirm this inference, we then generated virtual instances corresponding to the state $\epsilon=3$, 
where the Gaussian perturbation is entirely inside $\rdom$ as depicted in Fig.~\ref{fig:systems_circ}c.
We notice that all the virtual instances have low MSE. 
Therefore, we conclude that SymAE learned to differentiate the coherently-illuminated portion of the medium by the waves without 
the need for physics. 
This means, for seismic monitoring experiments, 
all the sources must coherently illuminate the time-lapse medium changes of interest.
}

\rev{
The experiment in Fig.~\ref{fig:intro_redatum} 
involves seismic-wave propagation in a complex 2-D structural model, which is commonly known as the
Marmousi model \cite{brougois1990marmousi} in exploration seismology. 
The structural complexities will lead to multipath propagation.
The P-wave velocity plots of this model for $\epsilon=1$ and $\epsilon=2$, with source-reciever geometry, are
in the Figs.~\ref{fig:systems_pm}a and \ref{fig:systems_pm}b, respectively.
The forcing term represents a random plane-wave source input at the bottom of the model.
The source wavelet $\src(t;\tau)$ is generated by convolving a Ricker wavelet whose dominant frequency is sampled from
$\{10,\-12.5,\-\ldots,20\}$\,Hz, with a random time series
(of $0.4\,$s duration)
sampled from a standard normal distribution.
In this case, the change in the medium parameters from $\epsilon=1$ to $\epsilon=2$ is coherently illuminated by all the plane-wave sources,
similar to the $\epsilon=3$ Gaussian perturbation of the previous example.
Therefore, the virtual measurements
after redatuming are expected to have low MSE, as confirmed by the results in Fig.~\ref{fig:systems_pm}.
\begin{figure*}
\centering
\includegraphics[width=\textwidth]{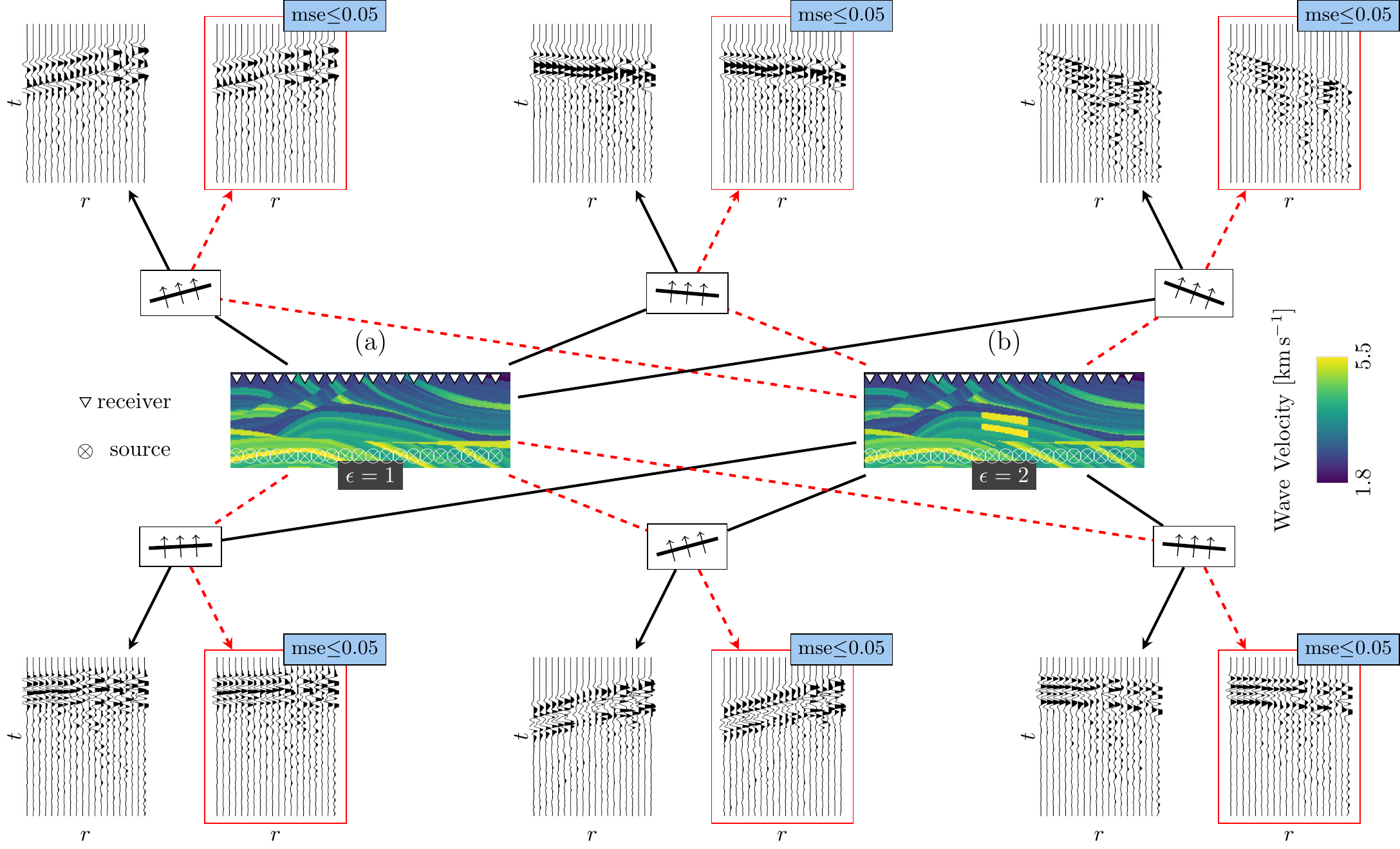}
\caption{
  \label{fig:systems_pm}
  Seismic Marmousi experiment of Fig.~\ref{fig:intro_redatum} where
  incident plane waves on the bottom of the medium undergo multipath propagation due to the complex inhomogeneities --- three original instances (solid arrows) from each state are 
  plotted reveal the complex wavefield.
  Each instance results from a planewave with a unique source wavelet and angle of arrival.
  Virtual instances (dashed arrows) generated after swapping the coherent code from (a) to (b) and vice versa have a low error, showcasing the success of deep redatuming with SymAE.
}
\end{figure*}
}

\paragraph*{Conclusions.}

We propose an autoencoder architecture for poorly controlled scientific experiments that produce an abundance of incompletely modeled measurements of a physical system. 
The autoencoder learns a data representation that disentangles the coherent information inherent to the physical state, from the nuisance modifications inherent to the experimental configuration, in a model-free fashion. 
Two ideas are critical: \begin{inparaenum} \item leveraging symmetry under reordering of the data instance in order to represent the coherent information in a first encoder, and \item stochastic regularization in order to prevent coherent information from being represented by the second encoder. \end{inparaenum}
As a result, the architecture can perform redatuming, i.e., the swapping of physics in order to create virtual measurements.

%



\begin{acknowledgments}
The authors thank TotalEnergies SE for their support.
PB is also funded via a start-up research grant from the Indian Institute of Science. 
PB thanks Girish from Indian Space Research Organisation and Shyama Narendranath from U R Rao Satellite Center for valuable discussions. 
\end{acknowledgments}


\section{Appendixes}

\subsection*{Algorithms}

These 
Tensorflow-style~\cite{Abadi2016}
algorithms
provide further details on the 
implementation of SymAE.
$\mathtt{ConvA}$ and $\mathtt{Conv}$ denote convolutional layers with and without $\mathtt{elu}$ activation, respectively. 
We used $\mathtt{distribute}$ to apply the same layer to each of the instances.
%



\begin{algorithm}
  \begin{algorithmic}[1]
	\caption{Keras-style algorithm for preparing coherent encoder $\encg$.}
	\label{alg:encg}
  \STATE $\Zg\gets \mathtt{distribute}(\mathtt{ConvA})(\X);\,\Zg\gets \mathtt{distribute}(\mathtt{ConvA})(\Zg);\, \Zg\gets \mathtt{distribute}(\mathtt{MaxPool})(\Zg)$ \hfill\COMMENT{$\encgone$}
  \STATE $\Zg\gets \mathtt{distribute}(\mathtt{ConvA})(\Zg);\,\Zg\gets \mathtt{distribute}(\mathtt{ConvA})(\Zg);\, \Zg\gets \mathtt{distribute}(\mathtt{MaxPool})(\Zg)$ \hfill\COMMENT{$\encgone  $}
  \STATE $\Zg\gets \mathtt{reduce\_mean}(\Zg,\text{axis}=1)$ \hfill\COMMENT{sum over instances}
  \STATE $\Zg\gets \mathtt{ConvA}(\Zg);\, \Zg\gets \mathtt{ConvA}(\Zg);\, \Zg\gets \mathtt{MaxPool}(\Zg)$ \hfill\COMMENT{$\encgtwo$}
  \STATE $\Zg\gets \mathtt{ConvA}(\Zg);\, \Zg\gets \mathtt{ConvA}(\Zg);\, \Zg\gets \mathtt{BatchNormalization}(\Zg)$ \hfill\COMMENT{$\encgtwo$}
  \STATE $\Zg\gets \mathtt{MaxPool}(\Zg);\,\Zg\gets \mathtt{Flatten}(\Zg);\,\Zg\gets \mathtt{Dense}(\Zg, \nZg)$ \hfill\COMMENT{$\encgtwo$}
  \STATE $\encg=\mathtt{Model}(\X, \Zg)$
  \end{algorithmic}
\end{algorithm}

\begin{algorithm}
  \begin{algorithmic}[1]
	\caption{Keras-style algorithm for preparing nuisance encoder $\encs$.}
	\label{alg:encs}
  \STATE $\Zs\gets \mathtt{distribute}(\mathtt{ConvA})(\X);\,\Zs\gets \mathtt{distribute}(\mathtt{ConvA})(\Zs);\, \Zs\gets \mathtt{distribute}(\mathtt{MaxPool})(\Zs)$
  \STATE $\Zs\gets \mathtt{distribute}(\mathtt{ConvA})(\Zs);\,\Zs\gets \mathtt{distribute}(\mathtt{ConvA})(\Zs);\, \Zs\gets \mathtt{distribute}(\mathtt{MaxPool})(\Zs)$
  \STATE $\Zs\gets \mathtt{distribute}(\mathtt{ConvA})(\Zs);\,\Zs\gets \mathtt{distribute}(\mathtt{ConvA})(\Zs);\, \Zs\gets \mathtt{distribute}(\mathtt{MaxPool})(\Zs)$
  \STATE $\Zs\gets \mathtt{distribute}(\mathtt{ConvA})(\Zs);\,\Zs\gets \mathtt{distribute}(\mathtt{ConvA})(\Zs)$
  \STATE $\Zs\gets \mathtt{distribute}(\mathtt{BatchNormalization})(\Zs)$
  \STATE $\Zs\gets \mathtt{distribute}(\mathtt{MaxPool})(\Zs);\,\Zs\gets \mathtt{distribute}(\mathtt{Flatten})(\Zs);\,\Zs\gets \mathtt{distribute}(\mathtt{Dense})(\Zs, \nZs)$
  \STATE $\encs=\mathtt{Model}(\X, \Zs)$
  \end{algorithmic}
\end{algorithm}

\begin{algorithm}
  \begin{algorithmic}[1]
  \caption{Fuse latent codes $\Zg$ and $\Zs$.}
	\label{alg:fuse}
  \STATE $\out{\Zs}\gets \mathtt{dropout}(\Zs, \alpha)$ \hfill\COMMENT{stochastic regularization}
  \STATE $\out{\Zg}=\mathtt{RepeatVector}(\ntau)({\Zg});\,\out{\X}\gets\mathtt{concatenate}([\out{\Zg},\out{\Zs}],\text{axis}=2)$ \hfill\COMMENT{distribute coherent code to each instance}
  \STATE $\hat{\X}\gets\mathtt{distribute}(\mathtt{Dense}(\nr\times n_t))(\hat{\X});\, \hat{\X}\gets\mathtt{distribute}(\mathtt{Reshape}(\nr,n_t,1))(\hat{\X})$
  \STATE $\hat{\X}\gets\mathtt{distribute}(\mathtt{ConvA})(\hat{\X});\,\hat{\X}\gets\mathtt{distribute}(\mathtt{ConvA})(\hat{\X});\,\hat{\X}\gets\mathtt{distribute}(\mathtt{ConvA})(\hat{\X})$
  \STATE $\hat{\X}\gets\mathtt{distribute}(\mathtt{BatchNormalization})(\hat{\X});\,\hat{\X}\gets\mathtt{distribute}(\mathtt{ConvA})(\hat{\X});$
  \STATE $\hat{\X}\gets\mathtt{distribute}(\mathtt{Conv})(\hat{\X});$\hfill\COMMENT{output datapoint}
  \end{algorithmic}
\end{algorithm}

\subsection*{Hyperparameters}
For a given application, the following hyperparameters need to be tuned: 
\begin{itemize}
\item filters and kernel sizes of the convolutional layers --- results do not strongly depend on these as long as the encoders and decoders have enough flexibility;
\item length of the coherent code $\nZg$ --- results do not strongly depend on this parameter;
\item the number of instances in each datapoint $\ntau$ --- we typically chose $\ntau=20$ and observed that results do not depend on 
this parameter as long as $\ntau>10$;
\item length of nuisance code $\nZs$ --- determined by the number of nuisance parameters;
\item and the dropout rate or noise strength $p$ added to the nuisance code --- stronger noise will lead to slower training, and weaker noise will make the coherent code dysfunctional.
\end{itemize}
Out of these, the parameters $\nZs$ and $p$ are crucial. They determine the balance between the noise and expressivity of the nuisance code. 
Increasing $\nZs$ proportional to the noise strength is essential as higher noise makes the nuisance code less expressive.
In our experiments, we typically chose $p=0.5$ and $\nZs$ as twice the number of nuisance parameters.


\bibliography{symae}

\clearpage
\end{document}